\documentclass[usenatbib]{mn2e}
\usepackage{graphicx}
\usepackage{amsmath}
\usepackage{amssymb}
\usepackage{blkarray}
\usepackage{color}
\usepackage{longtable}
\input psfig.sty

\font\hf = cmsl8 scaled \magstep 0

\title{Effect of dynamical interactions on integrated properties of globular clusters}

\author[Y. Zhuang et al.]
{Yulong~Zhuang\thanks{E-mail: ctanzhuangyl@Gmail.com}$^{1,2,3}$, Fenghui~Zhang$^{1,3}$, Peter~Anders$^{4}$,
 Zhifeng~Ruan$^{1,2,3}$, \and  Liantao~Cheng$^{1,2,3}$ and Xiaoyu~Kang$^{1,2,3}$\\
$^1$Yunnan Observatories, Chinese Academy of Sciences, Kunming, 650011, China \\
$^2$University of Chinese Acadamy of Sciences, Beijing, 100049, China\\
$^3$Key Laboratory of Structure and Evolution of Celestial Objects, Chinese Academy of Sciences, Kunming, 650011, China\\
$^4$Key Laboratory of Optical Astronomy, National Astronomical Observatories, Chinese Academy of Sciences, Beijing, 100012, China\\
}

\begin{document}

\date{\today}

\pagerange{\pageref{firstpage}--\pageref{lastpage}}

\pubyear{2008}

\maketitle

\label{firstpage}

\begin{abstract}
Globular Clusters (GCs) are generally treated as natural validators of simple stellar population (SSP) models. However, there are still some differences between real GCs and SSPs. In this work, we use a direct $N$-body simulation code {\hf Nbody6} to study the influences of dynamical interactions, metallicity and primordial binaries on Milky Way GCs' integrated properties. Our models start with $N=100,000$ stars, covering a metallicity range  $Z=0.0001 \sim 0.02$, a subset of our models contain primordial binaries, resulting in a binary fraction as currently observed at a model age of GCs.
 Stellar evolution and external tidal field representative for an average Milky Way GC are taken into consideration. The integrated colours and Lick indices are calculated using BaSeL and Bluered stellar spectral libraries separately.

By including dynamical interactions, our model clusters show integrated features (i.e., colours up to 0.01mag bluer, H$\beta$ up to 0.1\AA $ $ greater and [MgFe]$'$ 0.05\AA $ $ smaller) making the clusters appear slightly younger than the model clusters without dynamical interactions.
This effect is caused mainly by the preferential loss of low-mass stars which have a stronger contribution to redder passbands as well as different spectral
features compared to higher-mass stars. In addition, this effect is larger at lower metallicities. On the contrary, the incorporation of primordial binaries
reduces this effect.

\end{abstract}


\section{Introduction}
Globular clusters (GCs) are among the oldest surviving stellar objects in the universe \citep{2006ARA&A..44..193B}. As fossil records of galaxy formation and evolution, they contain a wealth of important information about their host galaxies such as their structure \citep{2006MNRAS.373..157B} and merger history \citep{2004ApJ...614L..29L}. Therefore it is very important to study the properties of GCs in order to derive properties of their host galaxies.

Stellar population synthesis (SPS) is a very useful tool to explore such stellar systems, especially for distant GCs. When using SPS to study GCs, GCs are usually treated as simple stellar population (SSP) systems, i.e., all stars are born at the same time and have identical chemical composition. In previous studies of SPS models, the standard SSP systems are assumed to have a continuous and static initial mass function (IMF), but real GCs are not exactly equivalent to such SSP systems (see \cite{2011A&A...525A.122P}, for a comparison between smooth IMFs and discrete IMFs). Among other effects, there are four differences between the idealized SSP systems and real clusters discussed here:

\begin{enumerate}
\item Stellar populations are affected by the dynamical evolution of the cluster, which leads to mass segregation and evaporation. In general, low-mass stars will escape from a cluster more easily due to energy equipartition caused by two-body relaxation \citep{1987degc.book.....S}. Thus, long term dynamical evolution will change the cluster's stellar mass function significantly \citep{1987degc.book.....S,2009A&A...502..817A}. Just as stellar evolutionary track mainly depends on its initial mass, the integrated spectrum of a cluster depends on its mass function (MF) at a given time. The dynamical evolution of GCs in a tidal field manifests itself as the change of GCs' MF, which is accompanied by different integrated features (related to spectrum, associated colours and spectral indices). After the appearance of the first red supergiants (RSG) and red giants (RGB)(i.e., after the first few Myr), the integrated flux in passbands redder than V band is dominated by RGB(RSG) and low-mass stars, while the main contributors to bluer passbands are stars near the main sequence turn-off (MSTO). As the cluster ages, the loss of low-mass stars will make the cluster bluer, on the other hand, with the MSTO shifts successively redwards and the loss of mid-MS stars, the flux received by bluer bands will also be reduced. However, the relative contribution of mid-MS stars and MSTO stars also strongly depends on the MF, and hence on the effects of dynamical evolution \citep{2009A&A...502..817A}.

\item Observed star clusters contain binaries. The evolution of binary stars can be very different from single stars (especially for close and/or interacting binaries), and binary interactions can also create some hot luminous objects such as blue stragglers (BSs) \citep{1994A&A...288..475P}, which are blue and luminous stars on an extension of the MS above the MSTO \citep{1953AJ.....58...61S}. The existence of these stars can also mimic the presence of younger SPs \citep{2004AJ....127.1513S,2008ApJ...689L..29C,2005MNRAS.364..503Z}. The effect of BSs is small on colours and spectra of GCs, but significant on Lick-indices \citep{2012MNRAS.421.1678Z}. Binaries also play a very important role in GCs dynamical evolution: through binary-binary and binary-single interactions even a small binary fraction in the core can be sufficient to heat the cluster core enough to postpone core-collapse significantly \citep{1992PASP..104..981H,2006MNRAS.368..677H,2010MNRAS.404.1835K}.

\item Giant stars (i.e., RGB, horizontal-branch stars (HB), asymptotic-giant-branch stars (AGB) and thermally pulsing asymptotic-giant-branch stars (TP-AGB)) as well as planetary nebulae (PN) have great influence on the integrated spectra, because some of them are very bright, and consequently have a great contribution to the GCs integrated features. However, real star clusters only have a finite number of stars and a discrete IMF. In a star cluster less massive than $10^{6}M_{\sun}$, the number of bright stars is so small, that they will cause stochastic fluctuations on the integrated properties of the cluster \citep{1977A&A....54..243B,2000ASPC..211..144M,2010A&A...521A..22F,2011A&A...525A.122P}.

\item Recent studies of globular clusters based on their chemical composition \citep{1971Obs....91..223O,1981ApJ...245L..79C,2011A&A...533A..69C} and the splitting of sequences in the colour-magnitude diagram \citep{2009IAUS..258..233P} have shown that many of them are not simple stellar populations being rather made of multiple generations.
\end{enumerate}

\cite{2007arXiv0704.3915B} used a parallel version of direct $N$-body code, {\hf Nbody6++}\citep{1999JCoAM.109..407S} and BaSeL 2.0 stellar spectral library (which can transform basic stellar parameters into spectra \citep{1998A&AS..130...65L}) to explore the effect of dynamical interactions on colours and mass-to-light ratio. They found integrated cluster colours will become bluer before the final stage of cluster dissolution because of energy equipartition. However, they did not include TP-AGB stars in their models. \cite{2008A&A...490..151K} presented an analytical description of the underlying stellar MF's evolution due to stellar evolution and dynamical dissolution. In their models the preferential loss of low-mass stars was approximated by a gradual increase in the lower mass limit of the stars presented in the cluster. They got the same result that clusters exhibiting the preferential loss of low-mass stars are bluer than standard models before the final 10\% of a cluster's lifetime. However, \cite{2009A&A...502..817A} reanalysed a set of $N$-body simulations of star clusters in a tidal field, fit the evolution of the stellar MF and parameterized it as a function of age and total cluster disruption time. They used this parameterization to compute {\hf GALEV} models \citep{2009MNRAS.396..462K} as a function of age, metallicity and total cluster disruption time, and found the dynamical interactions generally make the cluster redder and fainter, and clusters will appear to be older than they are. These previous results provide good comparisons with our work.

In this work, we use a set of $N$-body simulations to study the influence of dynamical evolution on GCs' stellar populations. We systematically discuss the evolution of integrated colours and Lick-indices of typical Milky Way GCs at different metallicities. In addition, the effect of primordial binaries is also taken into account.

This paper is structured as follows. We introduce our simulation method and models in Section 2. In Section 3 we present and analyze our results, which is followed by a discussion and conclusions in Section 4.

\section{SIMULATION PARAMETERS \& METHOD }
\begin{table*}
\caption{Three sets of our cluster models. $Z$ is metallicity, $N_0$ is the number of stars at the beginning of the simulation, $Nb_{0}$ is the number of primordial binaries. "Evolution" is the evolution method: 'D' indicates the cluster evolves with stellar evolution and dynamical interactions, while 'S' means the cluster evolves only with stellar evolution. }
\begin{tabular}{lccccccccc}
             & set $a)$     &&    &set $b)$      &&     &set $c)$       && \\ \hline
$Z$          &$N_0$   &$Nb_{0}$&evolution&$N_0$&$Nb_{0}$&evolution&$N_0$&$Nb_{0}$&evolution\\ \hline\hline
0.0001 & 100000 &0 &D  & 100000 &0 & S & 100000 &5000 &D   \\
0.0003 & 100000 &0 &D  & 100000 &0 & S &   &  & \\
0.001   & 100000 &0 &D  & 100000 &0 & S & 100000 &5000 &D   \\
0.004   & 100000 &0 &D  & 100000 &0 & S &   &  & \\
0.01     & 100000 &0 &D  & 100000 &0 & S & 100000 &5000 &D   \\
0.02     & 100000 &0 &D  & 100000 &0 & S &   &  &    \\ \hline
\end{tabular}
\label{Table:NBstart}
\end{table*}

We use the $N$-body code {\hf Nbody6} \citep{1999PASP..111.1333A,2000AdSAC..10..286A,2003gnbs.book.....A} to construct and evolve our models. We run the code's GPU version \citep{2012MNRAS.424..545N} on one NVIDIA GTX660 graphics cards. {\hf Nbody6} is a direct $N$-body simulation code using a 4$^{th}$ order integrator, containing a Hermite scheme, hierarchically blocked variable time steps and the Ahmad-Cohen neighbor scheme. Stellar evolution prescriptions are also included, based on \cite{2001MNRAS.323..630H} and \cite{2002DDA....33.0801H}, hence, in the simulations each particle represents one star  \citep{1999PASP..111.1333A,2000AdSAC..10..286A}. In this work, the stars evolve according to the stellar evolution prescriptions of \cite{2000MNRAS.315..543H}, which covers a range of metallicities from 0.0001 to 0.02 and includes all phases of stellar evolution. In our simulation, following a supernova event, when a new neutron star or black hole is detected, a velocity kick is randomly assigned from a Gaussian velocity kick distribution peaked at $190 $km/s \citep{1997MNRAS.291..569H}. BaSeL \citep{1998A&AS..130...65L} and Bluered \citep{2008A&A...485..823B} spectral libraries are used to calculate GC colours and Lick indices respectively.

\subsection{Parameters}

In our simulations, a \cite{2001MNRAS.322..231K} IMF is used within the stellar mass range 0.08 $-$ 50 $M_{\sun}$.  Similar to \cite{2012MNRAS.427..167S}, each of our models starts with a Plummer density profile, an initial half-mass radius of $r_{50\%} \approx 6.2$pc (which makes the initial half-mass relaxation time around 1.7Gyr) and evolves in a MW-like tidal field.
Our work did not take the early stage of the cluster into account, which shows strong variability. In general, the early gas ejection within a proto-cluster driven by stellar feedback will lead to an drastic mass loss \citep{2007MNRAS.380.1589B,2008MNRAS.384.1231B}. Besides, the intermediate-mass black hole (IMBH), which is thought to be harbored in the center of GCs by some groups \citep{2001AAS...198.5605M,2013MNRAS.435.3272T} and the multiple stellar populations found in some GCs \citep{2005ApJ...621..777P} are not included in our models. Our cluster models are similiar to the models presented by \cite{2012MNRAS.427..167S}, so their work provide a good comparison with the dynamical part of our work.

\subsubsection{Environment}
The properties of the host galaxy and cluster orbit play an important role in the evolution of clusters \citep{2014arXiv1402.2289M}. In this work, we use a MW-like tidal field which contains three components: a point-mass bulge, an extended smooth disk \citep{1975PASJ...27..533M} and a dark matter halo. In our models we use $M_b$=$10^{10}M_{\sun}$ and $M_d$=$5\times10^{10}M_{\sun}$ for bulge and disk mass, respectively \citep[see][]{2008ApJ...684.1143X}. The scale parameters for the Miyamoto disk are a$=$4 kpc, b$=$0.5 kpc.

We place our clusters at a typical orbit of Milky Way globular clusters with a galactocentric distance of $d_{gc} =$8.5 kpc. The orbit is about 25 deg inclined to the galactic disc. The apogalacticon is 9.5 kpc and perigalacticon 7.5 kpc with an orbital period of $\thicksim$ 0.2 Gyr.  The inclination results in a maximum height z $=$ 4 kpc, which is typical for many MW clusters \citep{1996A&A...313..119D,2012MNRAS.427..167S}.

\subsubsection{Primordial binaries}

Binary stars are common in our Milky Way. At least 50\% of field stars are in binary systems \citep{1983ARA&A..21..343A}. In young massive stellar populations, the binary fraction is close to one \citep{2007A&A...474...77K,2007ApJ...670..747K,2012MNRAS.422..794E}. For open clusters, dynamic models are usually evolved with binary fractions of $20\% \sim 50\%$ \citep{2005MNRAS.363..293H,2007MNRAS.374..344T}, as observations find higher binary fractions in these clusters \citep{1993AJ....106..181M}. Much lower binary fractions are observed in GCs: \cite{2012A&A...540A..16M} have measured the binary fractions of 59 GCs in the MW and commonly found values around $b_f \thickapprox$ 5\%. In our work, one set of our models take an primordial binary fraction of 5.3\%, initially containing 100000 stars, 95000 total systems, including 90000 single stars and 5000 binaries systems. These binaries follow an uniform distribution in the period range log($P$/day)=$0.2\sim 6$. Some of these primordial binary systems get disrupted early on, while new binaries may form during the cluster evolution due to two- or three-body interactions.
\subsubsection{Models}

We evolved three sets of models shown in Table~\ref{Table:NBstart}, sets a) and b) each consist of models with six different metallicities $Z=0.0001, 0.0003, 0.001, 0.004, 0.01 $ and $ 0.02$. Set a) evolves with dynamical interactions, while set b) does not take dynamical interaction into account as a comparison. Both sets a) and b) start with 100000 single stars and do not contain primordial binaries. Set c), uses an initial binary fraction of 5.3\% (95000 total systems, including 90000 single stars and 5000 binaries systems) as mentioned above. In general, GCs have ages around 12 Gyr \citep{2007ApJ...671..861H}, however when using Lick indices to deduce GC's age, some GCs are older than 14 Gyr \citep{2012MNRAS.421.1678Z,2008ApJ...689L..29C}, so all models are evolved to 16Gyr in this work.

\subsection{Analysis Method}

In this work, the following methods were applied to explore the evolution of stellar populations and integrated features.

\subsubsection{Current-IMF}

The cluster MF can not only be affected by dynamical interactions but also by stellar evolution. In order to describe the effect of preferential loss of low-mass stars, we define the cluster current-initial mass function (current-IMF): for a cluster at a given age, we derived the initial masses ($M_{\mathrm{zams}}$) of each remaining star (including remnants). These values provide the initial mass function of all stars which are in the cluster at this age. Compared with cluster MFs used in previous work, the current-IMF can show the change of stellar populations without the impact of stellar evolution.

\subsubsection{Integrated flux calculation}

As mentioned in Section 1. the underlying assumption in standard SSP models is the use of a continuous stellar IMF in clusters, but real clusters only have a finite number of stars. The number of bright stars is so small, that stochastic fluctuations in the photometric properties of the cluster are common \citep{2011Ap&SS.331....1W}. The discreteness of the IMF leads to bursts in magnitudes and colours of clusters at moments when bright stars appear and then die. The amplitude of the burst depends on the cluster mass and on the spectral range. The higher the cluster mass the smaller is the bias \citep{2011A&A...525A.122P}. \cite{2009AJ....138.1724P,2013ApJ...778..138A} identified caveats and quantified the fitting uncertainties associated with this standard procedure. They showed that this effect can yield highly unreliable fitting results.

In order to isolate the influence of dynamical interactions, for each model, the following method was applied to eliminate the impact of the stochastic numbers of giant stars (RGB stars, HB stars, AGB stars and TP-AGB stars) as well as of planetary nebulae. For MS stars, white dwarfs and neutron stars we calculated the spectra directly. For giant stars, similiar to \cite{2009A&A...502..817A}, we reconstructed a well-proportioned giant star distribution by using the derived current-IMF, then calculated its spectra. Finally we combined the two parts and got the colours and Lick indices. Therefore giant stars are redistributed idealizely by using the analysed current-IMF. Note, the method we used here did not change the systematic results of dynamic interactions. As the current-IMF is used, the idealized distribution of giant stars can still keep the effect of dynamic evolution, but the jumps in magnitudes and colours caused by the stochastic distribution of giant stars will be smoothed.

\section{Results}
\begin{table*}
\caption{Various parameters of our models (sets a-c) at 15 Gyr, $Z$ is metallicity, $N_{\mathrm{15Gyr}}$ is the number of stars, $Nb_{\mathrm{15Gyr}}$ is the number of binary systems, $M_{\mathrm{15Gyr}}/M_{\mathrm{0Gyr}}$ is the ratio of present mass to initial mass.}
\begin{tabular}{lccccccccc}
        & set $a)$     &&    &set $b)$      &&     &set $c)$       && \\ \hline
Z        &$N_{\mathrm{15Gyr}}$&$Nb_{\mathrm{15Gyr}}$&$M_{\mathrm{15Gyr}}/M_{\mathrm{0Gyr}}$&$N_{\mathrm{15Gyr}}$&$Nb_{\mathrm{15Gyr}}$&$M_{\mathrm{15Gyr}}/M_{\mathrm{0Gyr}}$&$N_{\mathrm{15Gyr}}$&$Nb_{\mathrm{15Gyr}}$&$M_{\mathrm{15Gyr}}/M_{\mathrm{0Gyr}}$\\ \hline\hline
0.0001 & 18556 &1 &0.1536 & 100000 &0 &0.7206& 21926 &1555 &0.1832  \\
0.0003 & 18503 &1 &0.1537 & 100000 &0 &0.7172&   &  & \\
0.001  & 19022 &2 &0.1579 & 100000 &0 &0.7096& 22881 &1642 &0.1908  \\
0.004  & 20855 &1 &0.1719 & 100000 &0 &0.7032&   &  & \\
0.01   & 22115 &2 &0.1868  & 100000 &0 &0.7005& 26145 &1733 &0.2238  \\
0.02   & 23882 &1 &0.2035 & 100000 &0 &0.7115&   &  &    \\ \hline
\end{tabular}
\label{Table:NBresult}
\end{table*}

\begin{figure}
\centering{
\includegraphics[ height=5cm,width=8cm,clip]{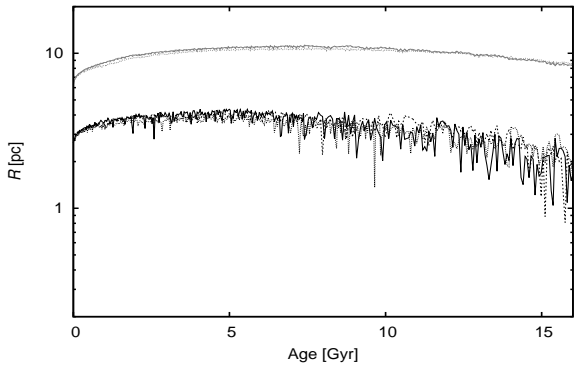}
}
\caption{The evolutions of $R_{50\%}$ (upper set) and $R_{\mathrm{core}}$ (lower set) of clusters with metalicity $Z=$ 0.0001 (solid lines), 0.001 (dashed lines) and 0.01 (dotted lines) in set a). }
\label{Fig:Radius}
\end{figure}

\begin{figure*}
\centering{
\includegraphics[ height=6cm,width=16.5cm,clip]{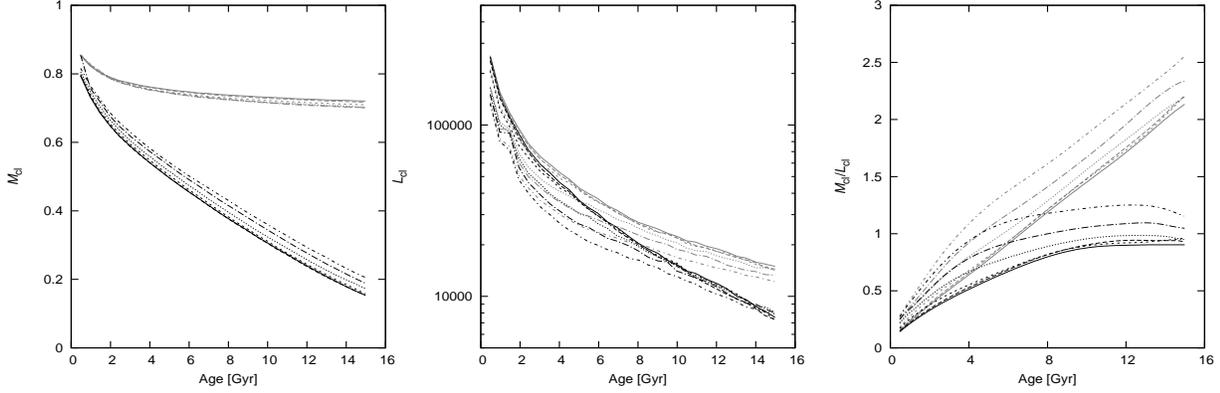}
}
\caption{The evolutions of cluster mass $M_{\mathrm{cl}}$ in units of cluster initial mass (left), luminosity $L_{\mathrm{cl}}$ in units of $L_{\sun}$ (middle) and mass-to-light ratio in units of $M_{\sun}/L_{\sun}$ (right) for set a) and set c). In each panel black lines denote the set a) models evolved with dynamical interactions and stellar evolution, grey lines denote the set b) models of clusters evolved without dynamical interactions (only with stellar evolution). Different lines imply different metallicities (solid: $Z=$0.0001, long dashed: $Z=$0.0003, short dashed: $Z=$0.001, dotted: $Z=$0.004, long dash-dotted: $Z=$0.01, short dash-dotted: $Z=$0.02). Note, the cluster properties are not plotted from zero-age for the sake of clarity.}
\label{Fig:DMLML}
\end{figure*}
\begin{figure}
\centering{
\includegraphics[ height=5cm,width=8cm,clip]{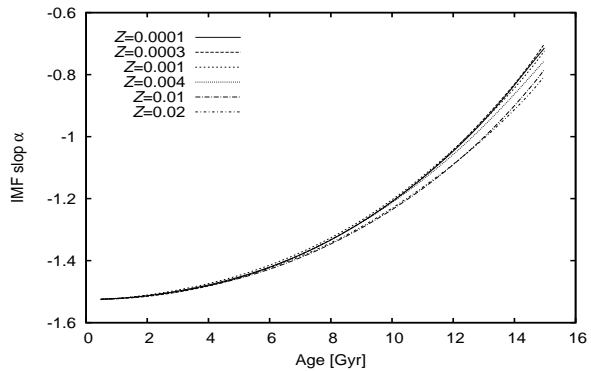}
}
\caption{The current-IMF slope between $0.25-0.7M_{\sun}$ for set a) as a function of cluster age. Line types have same meanings as in Fig.~\ref{Fig:DMLML}.}
\label{Fig:DIMF}
\end{figure}
\begin{figure*}
\centering{
\includegraphics[ height=6cm,width=16cm,clip]{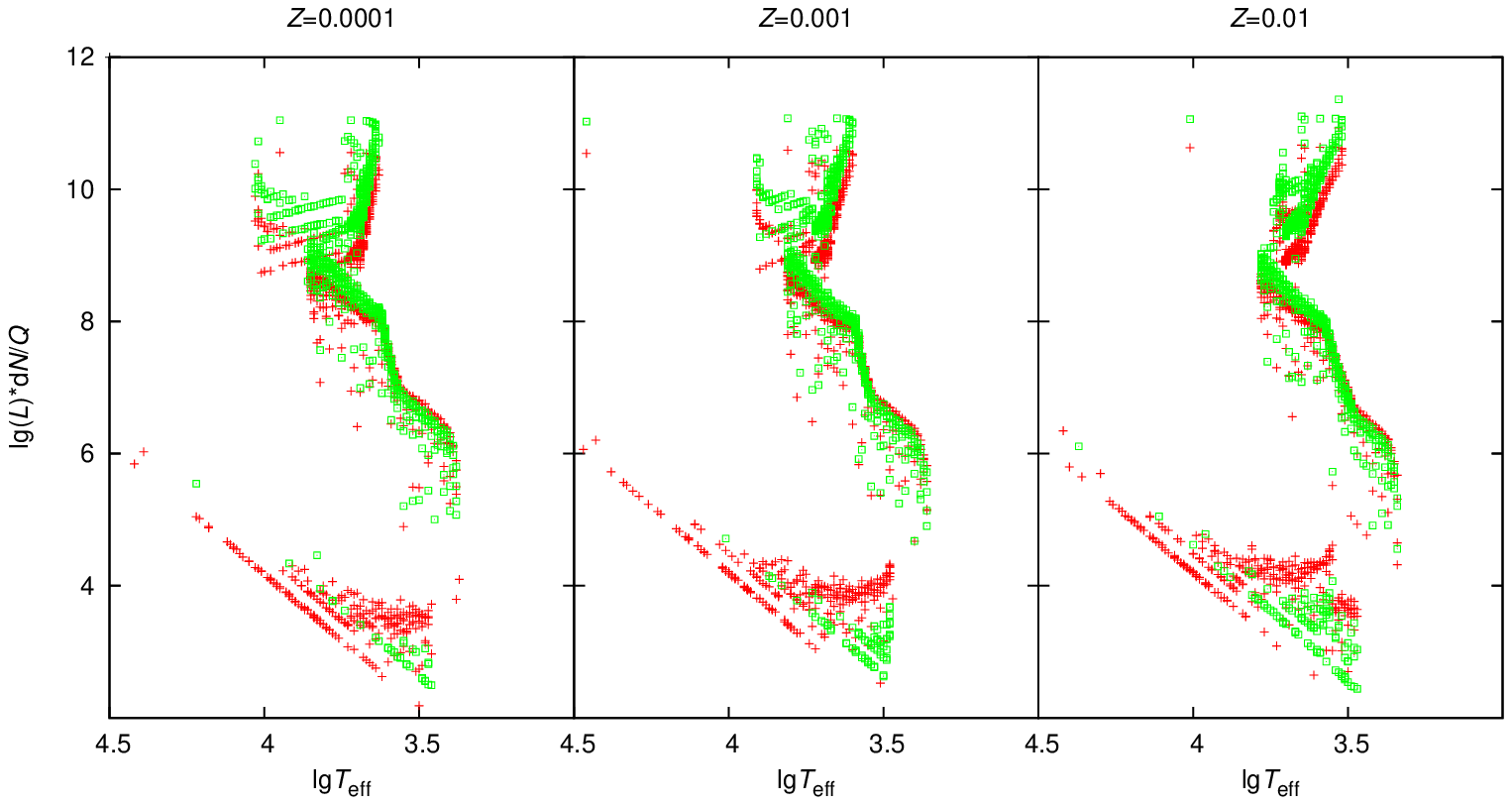}
\includegraphics[ height=6cm,width=16cm,clip]{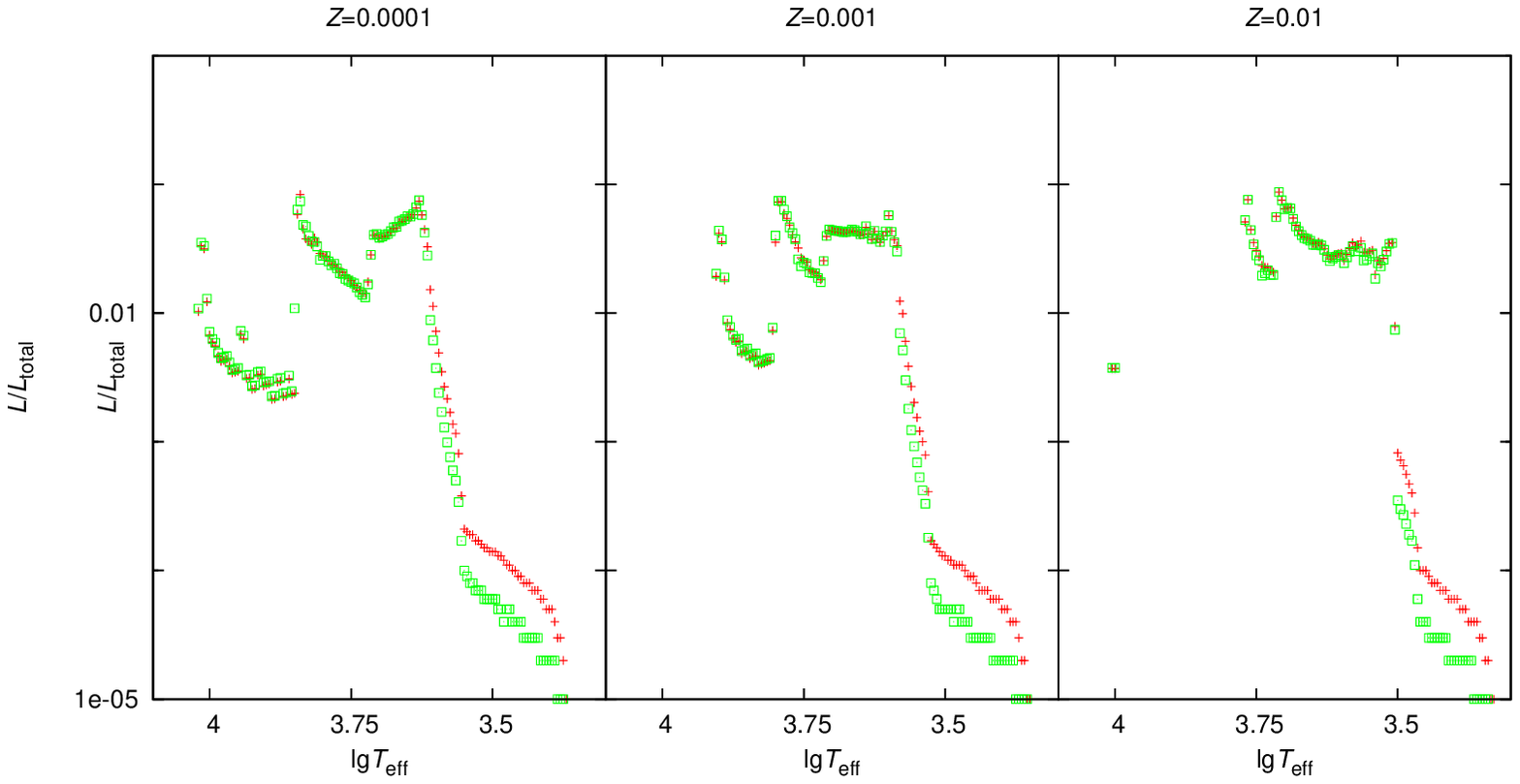}
}
\caption{Stellar populations of our cluster models at 12Gyr at different metallicities. Top panels are the modified HR diagram with $T_{\mathrm{eff}}$ (in units of K) against luminosity $L$ (in units of $L_{\sun}$) times $\mathrm{d}N/Q$, $Q$ is a normalisation constant. Bottom panels are $L$ versus $T_{\mathrm{eff}}$ distributions at three metallicities, the red crosses represent clusters evolve purely with stellar evolution, while green blocks are the models with dynamical evolution and stellar evolution. }
\label{Fig:Dste}
\end{figure*}

During our simulations, stars are lost through following channels:
\begin{enumerate}
\item Low-mass stars gain velocity through energy equipartition in numerous few-body encounters \citep{1971ApJ...164..399S};
\item Velocity kicks owing to the asymmetry of supernova explosions \citep{2004ApJ...601L.175F};
\item Tidal stripping by host galaxy.
\end{enumerate}
These effects can bring a star beyond the tidal boundary. Both channel (i) and (iii) will lead to the preferential loss of low-mass stars.
From the evolutions of $R_{50\%}$ and $R_{\mathrm{core}}$ of set a) showed in Fig.~\ref{Fig:Radius}, all clusters didn't reach core-collapse before 16Gyr and the metallicity didn't have an obvious effect on their size evolution.
The number and mass of remaining stars in our models at 15Gyr can be seen in Table~\ref{Table:NBresult}:
about $1/3$ of initial mass were lost solely due to stellar evolution (see set b)), $4/5$ of initial mass were lost when dynamical interactions are taken into account (see sets a) and c)). When evolving with dynamical interactions, clusters can retain more mass at high-$Z$ or with primordial binaries. The dynamic disruption time $T_{95\%}$ is around 21Gyr For set a), and around 25Gyr for set c), due to the binary systems delaying the core collapse. Here we use the
definition 'total cluster disruption time' \citep{2009A&A...502..817A} which is defined as the time when only $5\%$ of the initial cluster mass remains bound,
so our clusters didn't evolve to the final stage at 15Gyr.

The influence of dynamical interactions on stellar populations of GCs and integrated features will be discussed respectively. The effect of primordial binaries will be briefly discussed at the end of this section.

\subsection{Effect of dynamical interactions}

Fig.~\ref{Fig:DMLML} shows the present cluster mass $M_\mathrm{cl}$, luminosity $L_\mathrm{cl}$, and mass-to-light ratio $M_\mathrm{cl}/L_\mathrm{cl}$ of our models as a function of time. As shown in the left panel, due to the combined effect of dynamical interactions and stellar evolution, clusters in set a) (black lines) lost more mass than in set b) (grey lines), the mass loss only caused by stellar evolution. As mentioned before, the metallicity shows opposite effects on the mass loss rate between set a) and set b), due to the different time-scales of the formation of stellar remnants as a function of metallicity and their impact on cluster expansion \citep{2004MNRAS.355.1207H,2012MNRAS.425.2234D}.
Middle panel: like for mass evolution, due to dynamical interactions, the luminosity reduced more strongly in set a) than in set b). But the discrepancy is not as big as in mass evolution, because the dynamical interactions will cause preferential loss of low-mass stars and these kind of stars have a much higher mass-to-light ratio compared with massive stars.
With the high mass-loss rate dynamical interactions have a clear effect on the $M_\mathrm{cl}/L_\mathrm{cl}$ evolution: the preferential loss
of low-mass stars leads to a smaller $M_\mathrm{cl}/L_\mathrm{cl}$ (see the right panel), these results show good agreement with previous
work \citep[e.g.][]{2012MNRAS.427..167S}.

\subsubsection{Current-IMF and the modified Hertzsprung-Russel diagram}

Now we explore the influence of dynamical interactions on stellar populations. For set a), we use the present MF to derive the current-IMF for each age.
Fig.~\ref{Fig:DIMF} shows the evolution of the current-IMF slope $\alpha$ for stars in the range of $0.2M_{\sun} \leqslant M \leqslant 0.75M_{\sun}$ (as the MS turn-off mass in $Z$=0.0001 at 15Gyr is close to 0.75$M_{\sun}$) at different metallicities. The current-IMF slope becomes flatter as clusters ages because the number of low-mass stars is reduced significantly with time. It can also be seen that the change of $\alpha$ is greater in low-Z models. This is mainly due to the fact that metal-poor clusters suffer a higher mass loss rate, see Fig.~\ref{Fig:DMLML}.

The top panels of Fig.~\ref{Fig:Dste} are modified Hertzsprung-Russel(HR) diagrams which show the effective temperature against the luminosity times $\mathrm{d}N/Q$ ($\mathrm{d}N$ is the number of stars within $\mathrm{lg}L$ $\rightarrow$ $\mathrm{lg}L+\mathrm{dlg}L$ and $\mathrm{lg}T_{\mathrm{eff}}$ $\rightarrow$ $\mathrm{lg}T_{\mathrm{eff}}+\mathrm{dlg}T_{\mathrm{eff}}$; $Q$ is a normalisation constant proportional to the total number of stars of each model) of models with and without dynamical interactions at $T$=12Gyr. Hence these images show the relative proportion of stars in different $T_{\mathrm{eff}}$ and $L$ bins. For clarity, we only plot models in three metallicities $Z=$0.0001, 0.001 and 0.01. The vertical dispersion of horizontal branch (HB) and MS stars are caused by the scale of d$T_{\mathrm{eff}}$ and d$L$.
There is a prominent intersection between the dynamical interaction models and the pure stellar evolution ones (compared with models evolved only with stellar evolution, the dynamic models have smaller $\mathrm{lg}L\times\mathrm{d}N/Q$ at lower $T_{\mathrm{eff}}$, while have greater $\mathrm{lg}L\times\mathrm{d}N/Q$ at higher $T_{\mathrm{eff}}$), which is a natural result of the low-mass stars being removed. As low-mass stars have relatively low $T_{\mathrm{eff}}$, their loss will lead to a variation in some integrated properties. However, for giant stars, they do not show such tendency (their distributions do not have such intersection), because the mass of these stars is relatively large, so they haven't suffered significant differences in their current-IMF slopes before 12 Gyr. On the other hand these stars are only located in a very narrow mass range, thus the small change of current-IMF slope does not cause a strong influence on their populations.

The bottom panels of Fig.~\ref{Fig:Dste} show the luminosity contribution of different $T_{\mathrm{eff}}$ stars. Compared with models evolved with dynamical
interactions, the low $T_{\mathrm{eff}}$ stars will have a larger contribution to the total flux in models evolved only with stellar evolution. This effect is much stronger at low metallicity and mainly expressed in MS stars. On the other hand, giant stars do not show such effect and with the reduction of MS stars, they will have a relatively large contribution to the total flux.

\subsubsection{Colours}
\begin{figure*}
\centering{
\includegraphics[ height=6cm,width=16cm,clip]{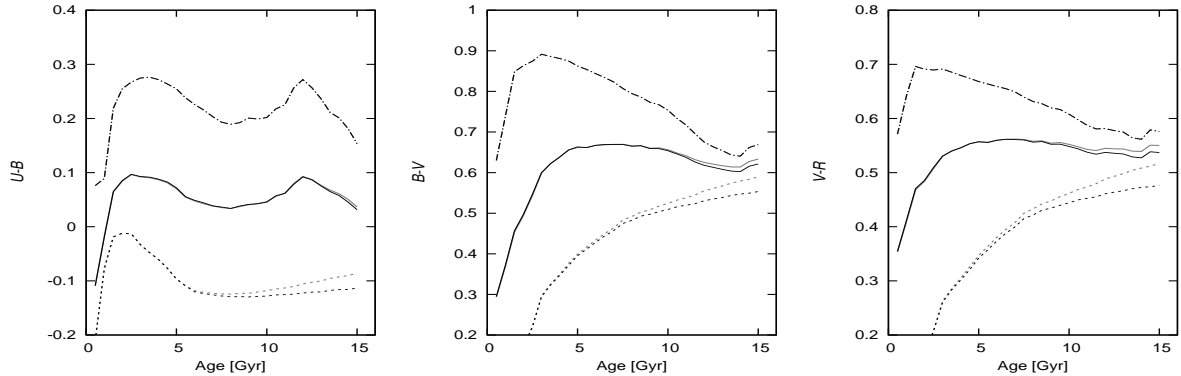}
}
\caption{Colour evolutions of different types of stars for sets a) and b) at $Z=$0.001 (right: $U-B$; middle: $B-V$; left:$V-R$). In each panel, dash-dotted lines are for the evolution of giant stars, dashed lines are the evolution of MS stars while solid lines are the colour evolution of an entire cluster. Black lines represent models with both dynamical interactions and stellar evolution, grey lines are models only with stellar evolution.   }
\label{Fig:DcolorA}
\end{figure*}
\begin{figure*}
\centering{
\includegraphics[ height=6cm,width=16cm,clip]{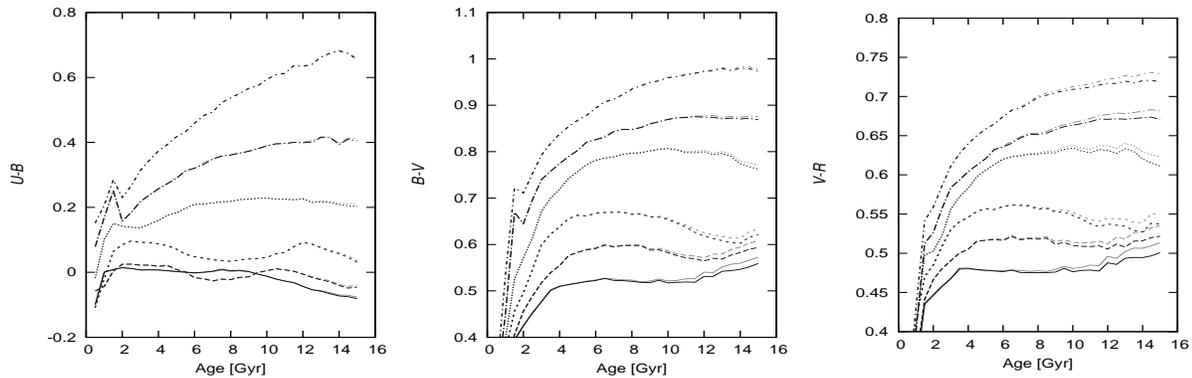}
}
\caption{Cluster integrated colour evolutions for sets a) and b) at six metallicities (right: $U-B$; middle: $B-V$; left: $V-R$). Black lines are clusters evolve with dynamical evolutions, while grey lines are clusters evolve only with stellar evolution. Line types have same meanings as in Fig.~\ref{Fig:DMLML}. }
\label{Fig:Dcolorall}
\end{figure*}
\begin{figure}
\centering{
\includegraphics[ height=5cm,width=9cm,clip]{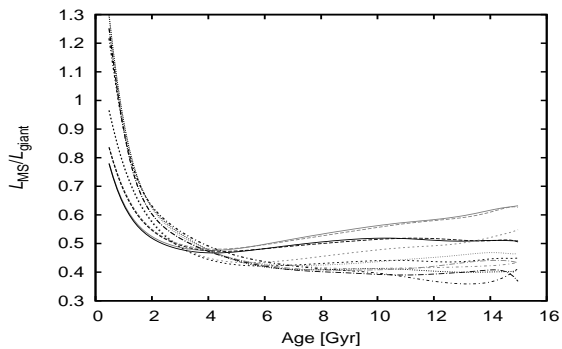}
}
\caption{The luminosity ratio between MS stars and giant stars at different metallicities for sets a) and b), for the sake of clarity, Bezier smoothing method was applied. Black lines represent clusters evolve with dynamical interactions and stellar evolution, grey lines are clusters evolve with pure stellar evolution. Line types have same meanings as in Fig.~\ref{Fig:DMLML}. }
\label{Fig:DLmsvsLrg}
\end{figure}

\begin{figure}
\centering{
\includegraphics[ height=5cm,width=9cm,clip]{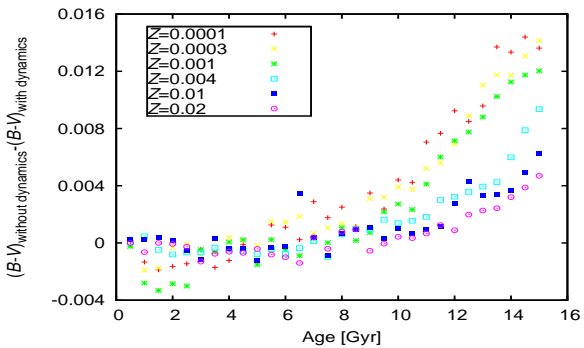}
}
\caption{$B-V$ divergence between set a) (evolve with dynamical interaction) and set b) (evolve only with stellar evolution) in six metallicities. }
\label{Fig:CAall}
\end{figure}

We transform stellar parameters ($T_{\mathrm{eff}}$, $\mathrm{log}g$ and $[\mathrm{M}/\mathrm{H}]$ ) into spectra and colours by using the stellar library BaSeL 2.0 \citep{1998A&AS..130...65L}. It covers the following fundamental parameters in the ranges listed
below: $T_{\mathrm{eff}}: 50,000\mathrm{K} \thicksim 2000\mathrm{K}$; $\mathrm{log}g: 5.5 \thicksim -1.02$, and $[\mathrm{M}/\mathrm{H}]: + 1.0 \thicksim -5.0$.
Linear interpolation is used to derive spectra in this work. For stars out of range, we use black-body spectra as an approximation.

Fig.~\ref{Fig:DcolorA} shows the influence of dynamical interactions on $U-B$, $B-V$, $V-R$ colour evolutions derived from different type of stars for sets a)
and sets b). For MS stars, with the loss of low $T_{\mathrm{eff}}$ stars (see 3.1.1), their flux contribution in redder bands is reduced. As a result, dynamical
interactions caused a difference in colours, which become bluer and the divergence increases with time (up to $0.02$mag at 15Gyr). However, the dynamical interactions did not show an obvious impact on colours of giant stars. The cluster's integrated colour is the combination of the contribution of these two type of stars and stellar remnants (we do not plot the colour evolution of stellar remnants alone, as they have little contribution to the total flux) which exhibits a smaller difference.
Fig.~\ref{Fig:Dcolorall} displays the evolution of integrated colours $U-B$, $B-V$ and $V-R$ for sets a) and b). In general, clusters will become redder as they ages, because of the shorter life time of high-mass stars. With dynamical interactions, more low-mass stars are lost, this results in clusters bluer than their no-dynamic interaction counterparts, and this effect will become greater with time. However, as showed in Fig.~\ref{Fig:DcolorA} giant stars are less influenced by this effect, the integrated colour divergence is relatively small (up to 0.01mag). In addition, the change of stellar population mainly happen in low-mass stars. These stars also have low $T_{\mathrm{eff}}$, so the colour divergence is larger in redder bandpasses. As shown in Fig.~\ref{Fig:Dcolorall}, the $U-B$ colour does not show differences as large as these of $B-V$ or $V-R$.

As the effect of dynamical interactions on integrated colours also depends on the ratio of MS stars' contribution in the total flux, the color discrepancy varies with metallicity. Fig.~\ref{Fig:DLmsvsLrg} illustrates the luminosity ratio between MS stars and giant stars $L_{\mathrm{MS}}/L_{\mathrm{giant}}$ for sets a) and b) at different metallicities. After 7 Gyr, in metal-poor cases the proportion of MS stars' luminosity becomes larger than in metal-rich cases for both sets a) and b). It is mainly because MS stars are much dimmer at high metallicities with a given initial mass. So $L_{\mathrm{MS}}/L_{\mathrm{giant}}$ is smaller at high metallicities and the dynamical interaction exerts less influence on the integrated colours. As the integrated $B-V$ divergence shown in Fig.~\ref{Fig:CAall}, the effect of dynamical interactions on $B-V$ becomes larger as metallicity is reduced.

\subsubsection{Lick indices}
\begin{figure*}
\centering{
\includegraphics[ height=7.5cm,width=16cm,clip]{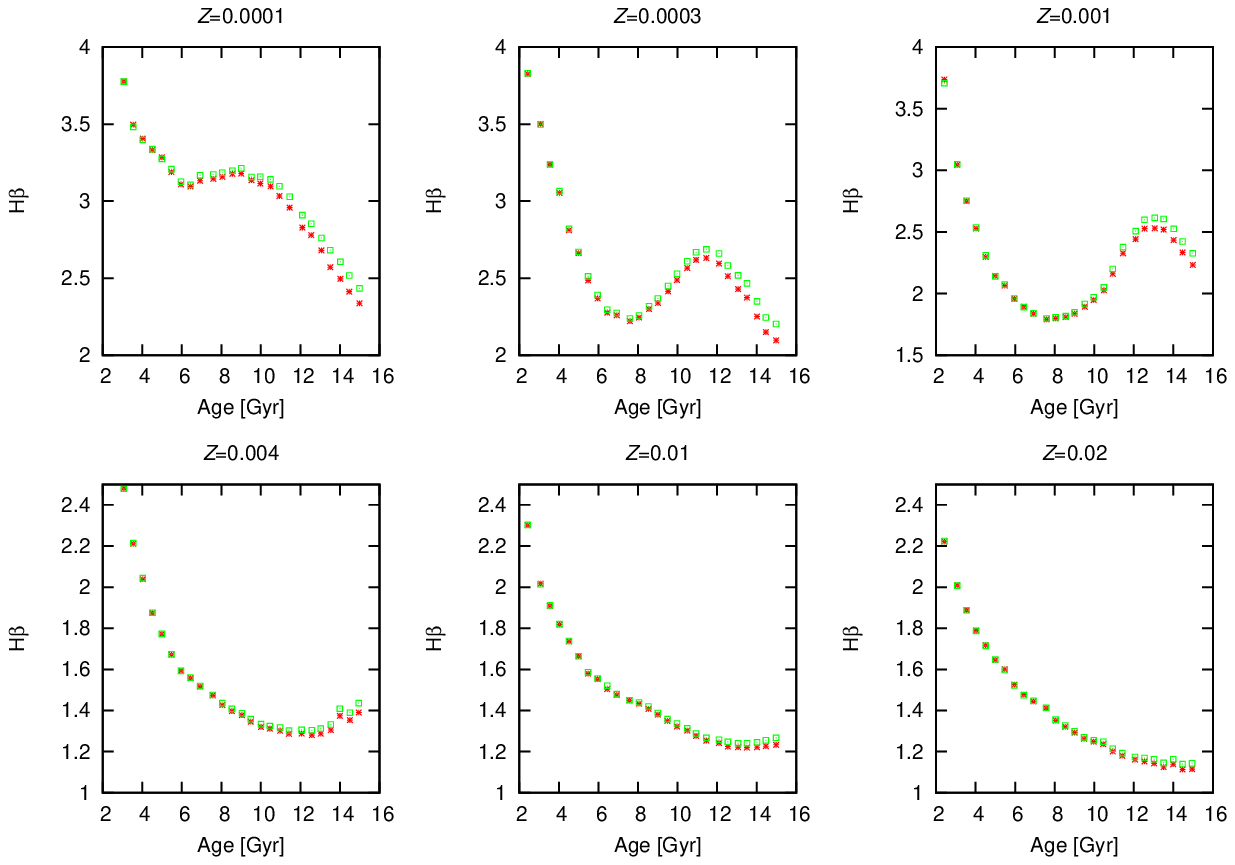}
\includegraphics[ height=7.5cm,width=16cm,clip]{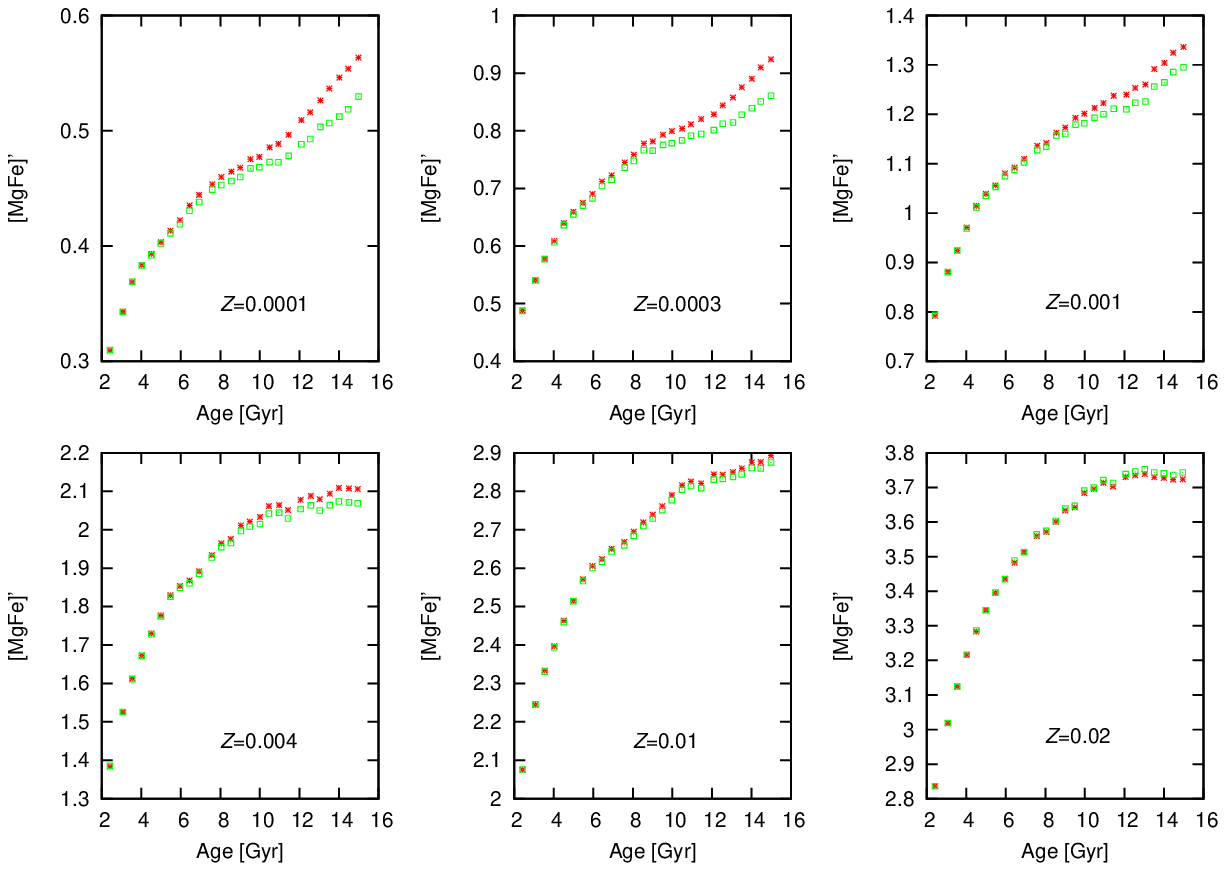}
}
\caption{Evolution of lick indices of our cluster models for sets a) and b) with six metallicities. Top six panels are the H$\beta$ index. Bottom six panels are [MgFe]$'$ index. In each panel, red crosses represent clusters evolve purely with stellar evolution, while green rectangles are clusters evolve with dynamical interactions and stellar evolution. }
\label{Fig:DLick}
\end{figure*}

Due to the existence of the age-metallicity degeneracy \citep{1994ApJS...94..687W,2002MNRAS.330..547T,2005MNRAS.362...41G}, it is not reliable to only use colour to interpret the stellar content. In contrast, some spectral features are sensitive to age or metallicity, i.e., the spectral indices. They can be used to break the age-metallicity degeneracy of stellar population systems. Thus, Lick indices have been widely used to determine parameters of stellar populations \citep{1994ApJS...94..687W,1998ApJS..116....1T}.

\begin{figure*}
\centering{
\includegraphics[ height=5cm,width=16cm,clip]{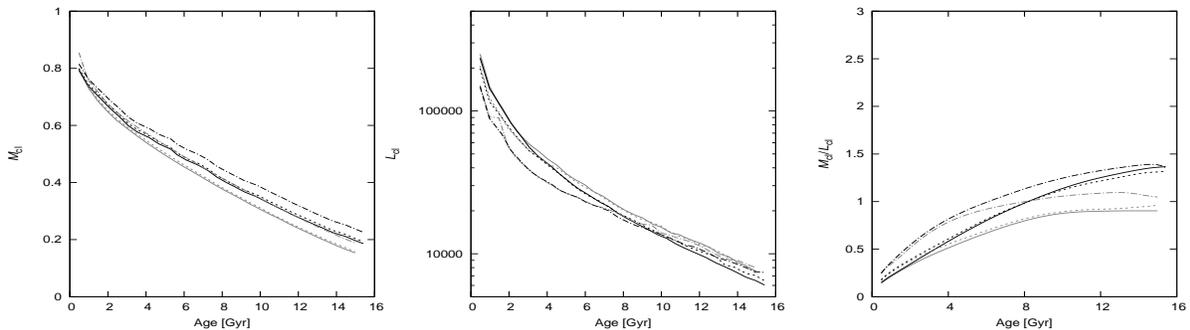}
}
\caption{Mass evolution in units of cluster initial mass (left panel), luminosity evolution in units of $L_{\sun}$ (middle panel) and the $M_{\mathrm{cl}}/L_{\mathrm{cl}}$ evolution in units of $M_{\sun}/L_{\sun}$(right panel) for sets a) and c) at 3 metallicities (solid lines: $Z=0.0001$, dashed lines: $Z=0.001$, dash-dotted lines: $Z=0.01$). Black lines indicate models with primordial binaries while grey lines do not include primordial binaries. }
\label{Fig:BMLML}
\end{figure*}

\begin{figure*}
\centering{
\includegraphics[ height=5.5cm,width=16cm,clip]{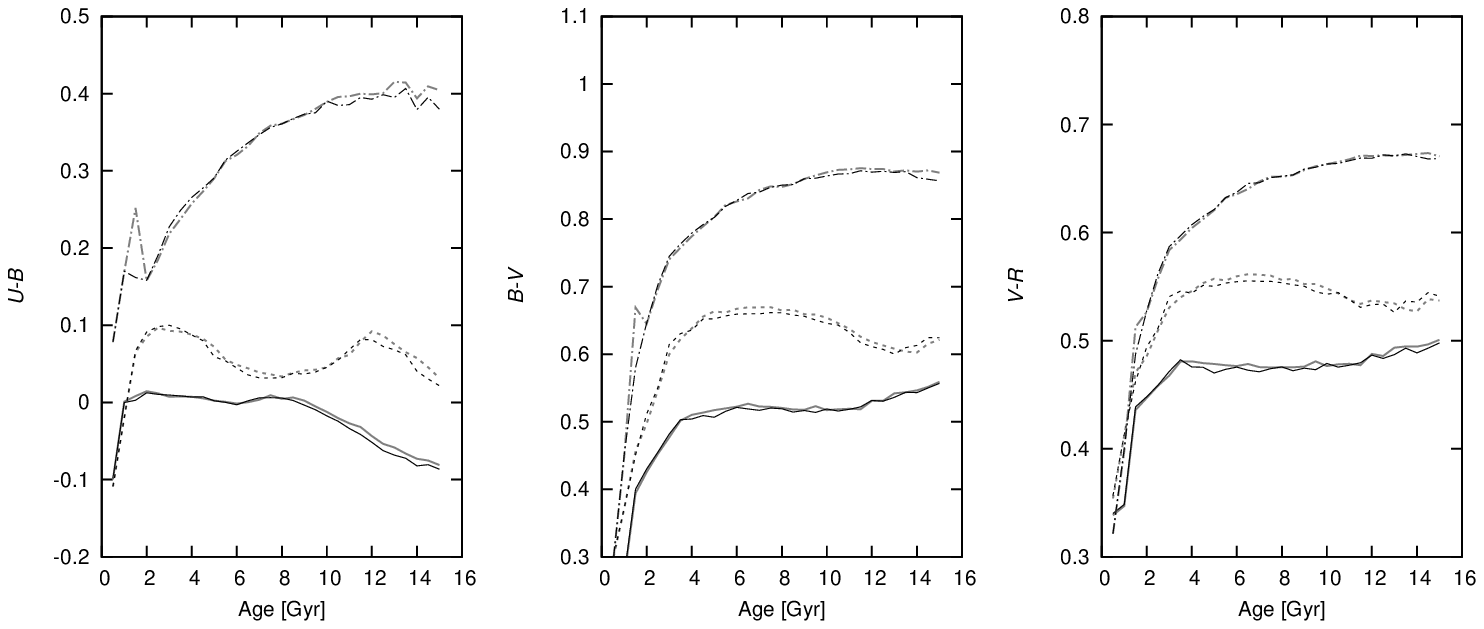}
}
\caption{Colour evolutions for sets a) and c) at 3 metallicities (Right: $U-B$; middle: $B-V$; left: $V-R$). Black lines are models including primordial binaries while grey lines do not include them. Line types have the same meanings as in Fig.~\ref{Fig:BMLML}. }
\label{Fig:Bcolorall}
\end{figure*}

In 1994, Worthey defined 21 Lick/IDS absorption line indices, based on a sample of 460 stars observed by a group at Lick Observatory. Then 4 new indices was
defined by \cite{1997ApJS..111..377W}, and \cite{1998yCat..21110377W} redefined the first 21 indices. These 25 indices have become the most widely used indices.

\begin{table*}
\caption{Definition of 8 Lick-indices used in this work.}
\begin{tabular}{lcccc}
 Name                 & Index band (\AA)       & blue continuum (\AA)    & red continuum (\AA)     &units \\ \hline
 $H_{\beta}$          & 4847.875-4876.625 &  4827.875-4847.875 &  4876.625-4891.625  & \AA \\
 Mgb                  & 5160.125-5192.625 &  5142.625-5161.375 &  5191.375-5206.375  & \AA \\
 Fe5270               & 5245.650-5285.650 &  5233.150-5248.150 &  5285.650-5318.150  & \AA \\
 Fe5335               & 5312.125-5352.125 &  5304.625-5315.875 &  5353.375-5363.375  & \AA \\
 $\mathrm{H}\delta_A$ & 4083.500-4122.250 &  4041.600-4079.750 &  4128.500-4161.000  & \AA \\
 $\mathrm{H}\gamma_A$ & 4319.750-4363.500 &  4283.500-4319.750 &  4367.250-4419.750  & \AA \\
 $\mathrm{H}\delta_F$ & 4091.000-4112.250 &  4057.250-4088.500 &  4114.750-4137.250  & \AA \\
 $\mathrm{H}\gamma_F$ & 4331.250-4352.250 &  4283.500-4319.750 &  4354.750-4384.750  & \AA \\  \hline
\end{tabular}
\label{Table:lick}
\end{table*}

In this work, we also explore the effect of dynamical interactions on the evolution of Lick indices, we are mainly focusing on 6 widely used indices: 5 age-sensitive Balmer indices ($\mathrm{H}\beta, \mathrm{H}\gamma_A, \mathrm{H}\gamma_F, \mathrm{H}\delta_A$ and  $\mathrm{H}\delta_F$), and one metal-sensitive index [MgFe]$'$, which is a combination of 3 metal-sensitive indices Mgb, Fe5270 and Fe5335:

$$ [\mathrm{MgFe}]'=[\mathrm{Mgb}(0.72 \times \mathrm{Fe}5270 + 0.28 \times \mathrm{Fe}5335)]^{1/2} $$

As $\alpha$ elements ($Ne$, $Mg$, $Si$, $S$, $Ar$, $Ca$ and $Ti$) have counteracting impact on Mgb, Fe$5270$ and Fe$5335$. This certifies that the [MgFe]$'$ index is independent of $\alpha$ elements abundance ($\alpha$/Fe) and can be considered as a good tracer of metallicity \citep{2003MNRAS.339..897T}. The definition of these indices are showed in Table~\ref{Table:lick}.

We use the high resolution ($R\geq500000$) Bluered spectra lib \citep{2008A&A...485..823B} to calculate spectral indices. Again, for stars beyond the library's parameter range we use a black-body spectrum as an approximation.

Fig.~\ref{Fig:DLick} top six panels plot $\mathrm{H}\beta$ evolutions with the effect of dynamical interactions in 6 different metallicities. H$\beta$ becomes smaller as cluster ages, because it is sensitive to the $T_{\mathrm{eff}}$. Due to the preferential loss of low $T_{\mathrm{eff}}$ stars, $\mathrm{H}\beta$ become slightly larger (up to 0.1\AA) than that of models evolved purely with stellar evolution. The upturn after 8Gyr in low-metallicity models is caused by the contribution of hot blue horizontal branch stars. The same trend is shown in the evolution of $\mathrm{H}\gamma_A, \mathrm{H}\delta_F, \mathrm{H}\gamma_A$ and $\mathrm{H}\delta_F$, for the sake of paper's size, we do not provide similar plots as they are very much like Fig.~\ref{Fig:DLick}.
The bottom six panels show the evolution of the combined index [MgFe]$'$. This index is metallicity-sensitive but also affected by $T_{\mathrm{eff}}$. In all of
our models this index tends to increase as cluster ages. Due to the preferential loss of low $T_{\mathrm{eff}}$ stars the index becomes smaller (up to 0.05\AA)
than the model evolved without dynamical interactions. For each indices, similar to the colour evolution, the divergence begins at early time and becomes larger at metal-poor models.

\subsection{Effect of primordial binaries}

\begin{figure*}
\centering{
\includegraphics[ height=8cm,width=16cm,clip]{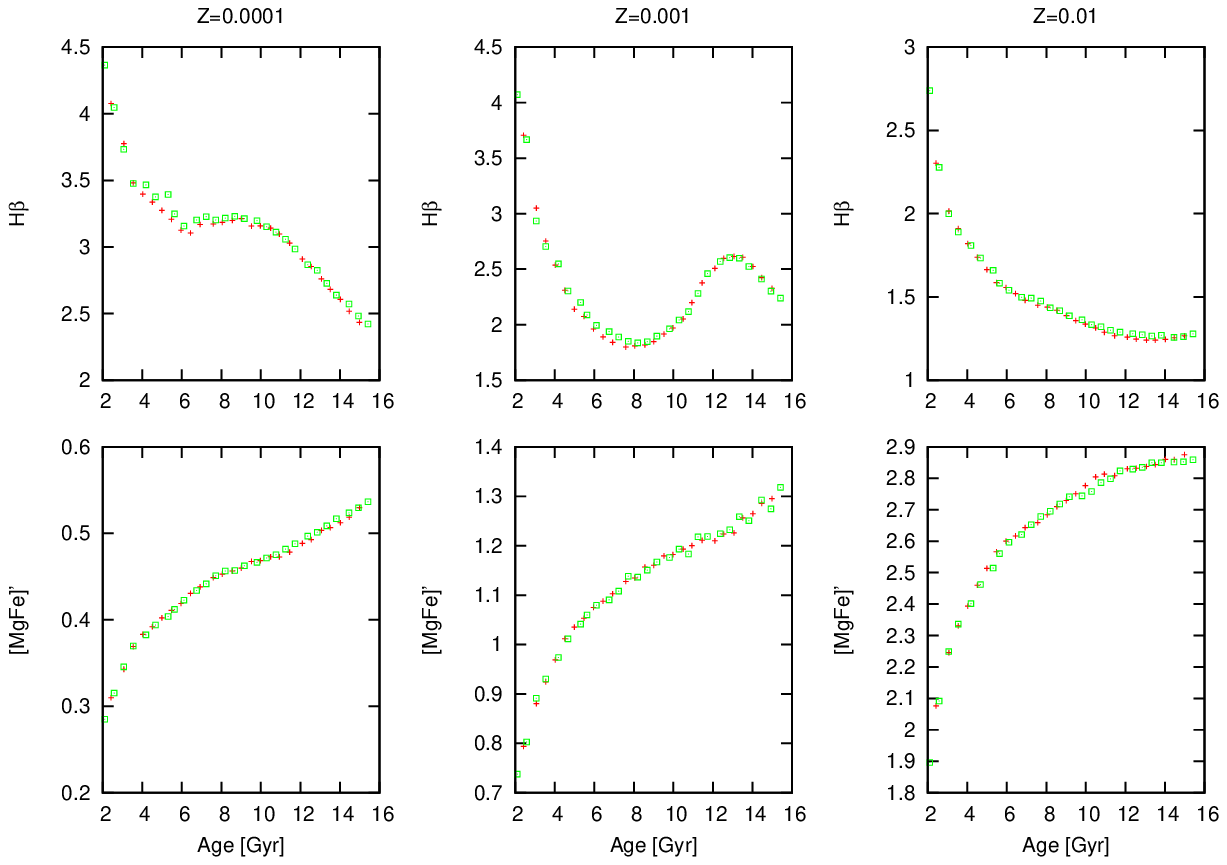}
}
\caption{The evolutions of Lick indices of our cluster models at three metallicities. Top panels are the evolutions of H$\beta$ index, bottom panels are [MgFe]$'$ indices. The red crosses represent clusters evolve without primordial binaries, while green rectangles are clusters with primordial binaries. }
\label{Fig:BLick}
\end{figure*}

In previous studies, some researchers have investigated the effect of primordial binaries on observations \citep{2008A&A...492..685D}. \cite{2005MNRAS.363..293H} have included binary interactions in the {\hf Nbody4} code \citep{1999PASP..111.1333A}. \cite{1999JCoAM.109..407S} and \cite{2001MNRAS.321..199P} have included binaries in the {\hf Nbody6++} and {\hf STARLAB} codes, respectively. \cite{2012MNRAS.421.1678Z} used the binary stellar population (BSP) models with a binary fraction of 50\%, but didn't take dynamical interactions into account. They found the five Balmer indices of BSP model are greater by 0.15\AA $ $ than those of SSP model.

In this work we explored the effect of primordial binaries with dynamical interactions, see set c) in Table~\ref{Table:NBstart}. The number and mass of  remaining stars at 15 Gyr are shown in Table~\ref{Table:NBresult}, For all three metallicities, the binary fraction of initially 5.3\% slightly increases to around 6\% $\sim$ 7\% at 15Gyr, which is a good reproduction of \cite{2012MNRAS.427..167S}. Compared with model a) clusters can keep more stars, because of the existence of primordial binaries can slow down the evaporation rate and the preferential loss of low-mass stars \citep{2007A&A...474...77K}. In this way the effect of dynamical interactions is reduced. Fig.~\ref{Fig:BMLML} is the evolution of cluster mass $M_{\mathrm{cl}}$, luminosity $L_{\mathrm{cl}}$ and mass-to-light ratio $M_\mathrm{cl}/L_{\mathrm{cl}}$ for sets a) and c). Obviously, with primordial binaries clusters can keep more mass, the luminosity is a bit brighter compared with the simple stellar population model, and the divergence in luminosity is smaller than that in mass, thus, the $M_\mathrm{cl}/L_{\mathrm{cl}}$ is larger than set a) and become higher as cluster ages. All these parameters show the same trend with the variation of metallicity as the models without primordial binaries.

The primordial binaries have two different effects on the integrated properties.

\begin{enumerate}
\item The existence of primordial binaries can slow down the evaporation rate, hence the cluster seems to be redder than the model without primordial binaries.

\item The existence of primordial binaries will lead to an increased number of hot luminous objects (such as BSs), which will make the cluster bluer.
\end{enumerate}

Fig.~\ref{Fig:Bcolorall} shows the colour evolution with the effect of the preferential loss of low-mass stars and the effect of primordial binaries. Like
Fig.~\ref{Fig:DcolorA} the evolution of the $U-B$, $B-V$ and $V-R$ colours at three metallicities, colour indices turn redder as the cluster ages, with the primordial binaries the cluster become slightly bluer in $U-B$, but in $B-V$ and $V-R$ the existence of primordial binaries did not result in obvious
changes due to the high $T_{\mathrm{eff}}$ objects mainly contribute in blue bandpasses. The effect of the slowing down of the evaporation rate can not be seen in the colour evolution, because of the very low luminosity of these additional remained stars.

In Fig.~\ref{Fig:BLick} we plot the evolution of $\mathrm{H}\beta$ and [MgFe]$'$ for sets a) and c) at $Z=$ 0.0001, 0.001 and 0.01. Models with 5.3\% primordial
binaries did not show any significant divergence with model a), probably because the effect of primordial binaries in GC's current-IMF is not of the same order as the effect of dynamical interaction and mainly act on low-mass range. These stars didn't play an important role for Lick indices. The comparably larger $\mathrm{H}\beta$ at some ages, especially between 4Gyr to 6Gyr, are caused by the existence of blue stragglers.

\section{summary and discussions}
We have used $N$-body simulations to explore the influences of dynamical interactions and primordial binaries on the evolution of galactic globular clusters with different metallicities $Z=$0.0001 $\sim$ 0.02. We focussed on their impacts on cluster mass, current-IMF, mass-to-light ratio, integrated colours and Lick indices. Our models based on the direct dynamical evolution code {\hf Nbody6} \citep{1999PASP..111.1333A}, and SSE and BSE stellar evolution code \citep{2001MNRAS.323..630H,2002DDA....33.0801H}.

Our results show that the dynamical interactions can change the stellar population of globular cluster by its selective effect on preferential loss of low-mass
stars (low-mass MS stars and stellar remnants). As these stars are rather faint, their loss will lead to a significant reduction of a cluster's mass-to-light
ratio, due to the significant mass loss and small luminosity reduction. This effect is larger for low metallicity clusters. Regarding colour evolutions: as the low MS stars have a relatively low effective temperature, their escape will make the cluster bluer which is in agreement with \cite{2008A&A...490..151K} and \cite{2007arXiv0704.3915B}.
In addition, this effect is found to be greater at low metallicities and redder colours ($B-V$ and $V-R$). In terms of Lick indices, the loss of
low $T_{\mathrm{eff}}$ stars causes larger Balmer indices and smaller [MgFe]$'$ index compared with models without dynamical interactions. Also, this effect is greater for low metallicities, due to MS stars contribute less luminosity at higher metallicities. The existence of primordial binaries reduces the mass loss rate, and consequently prevents the preferential loss of low-mass stars. Therefore, the cluster mass is larger than for the model without primordial binaries, so are luminosity and mass-to-light ratio. For colours, the effect is different. As the primordial binaries will produce some hot luminous objects, the cluster will consequently become slightly bluer in $U-B$. On a separate note, the Lick indices didn't show strong changes.

In this work, all clusters didn't suffer a core collapse before 16Gyr, so all of our results are based on pre-core collapse clusters. As core collapse will accelerate the loss of stars, the effect of dynamical interactions will be greater on clusters' integrated properties. On the other hand, as we focused on the influence of dynamical interactions, we eliminated the stochastic mass distribution of giant stars as it could lead to unreliable fitting results, so this effect was not included in our model. Besides, recently, there are some new evidences suggest the MW halo mass is under $10^{12}M_{\sun}$ \citep{2012ApJ...761...98K,2014ApJ...794...59K}, It may also bring some interesting effects. All data of our work will be uploaded to our website (www1.ynao.ac.cn/$\sim$zhangfh)

\section*{acknowledgements}
We thank the referee for the careful review and instructive comments that have improved the quality of this manuscript.
This work was funded by the Chinese Natural Science Foundation (Grant Nos 11273053 \& 11403092).
Many thanks to Thijs Kouwenhoven, Sverre Aarseth, Dengkai Jiang, Yu Zhang, Anna Sippel, Yan Gao, Jarrod Hurley and Hao Li for their helpful discussions and comments.
\bibliographystyle{mn2e}
\bibliography{mybib}

\bsp
\label{lastpage}
\end{document}